# Entropy Production Rate is Maximized in Non-Contractile Actomyosin

Daniel S. Seara[1,3], Vikrant Yadav[2,3+], Ian Linsmeier[2,3+], A. Pasha Tabatabai[2,3], Patrick W. Oakes[4], S.M. Ali Tabei[5], Shiladitya Banerjee[6*], and Michael P. Murrell[1,2,3*]

[1] Department of Physics, Yale University, 217 Prospect Street, New Haven, Connecticut 06511, USA

[2] Department of Biomedical Engineering, Yale University, 55 Prospect Street, New Haven, Connecticut 06511, USA

[3] Systems Biology Institute, Yale University, 850 West Campus Drive, West Haven, Connecticut 06516, USA

[4] Department of Physics & Astronomy, and Department of Biology, University of Rochester, Rochester, New York 14627, USA

[5] Physics Department, University of Northern Iowa, Cedar Falls, Iowa 50614, USA

[6] Department of Physics and Astronomy, Institute for the Physics of Living Systems, University College London, Gower Street, London WC1E 6BT, UK

*=corresponding author

+These authors contributed equally

ABSTRACT

The actin cytoskeleton is an active semi-flexible polymer network whose non-equilibrium properties coordinate both stable and contractile behaviors to maintain or change cell shape. While myosin motors drive the actin cytoskeleton out-of-equilibrium, the role of myosin-driven active stresses in the accumulation and dissipation of mechanical energy is unclear. To investigate this, we synthesize an actomyosin material *in vitro* whose active stress content can tune the network from stable to contractile. Each increment in activity determines a characteristic spectrum of actin filament fluctuations which is used to calculate the total mechanical work and the production of entropy in the material. We find that the balance of work and entropy does not increase monotonically and, surprisingly, the entropy production rate is maximized in the non-contractile, stable state. Our study provides evidence that the origins of system entropy production and activity-dependent dissipation arise from disorder in the molecular interactions between actin and myosin.

**INTRODUCTION**

The eukaryotic cytoskeleton is an active, viscoelastic material that exhibits a wide range of dynamic responses to both its internal and external environment[1]. This includes polarizing contractile flows during embryonic development [2] and cell division in the adult [3,4]. By contrast, there are dynamic steady states, including ratcheting motions in the *Drosophila* wing [5], excitable wave motion in the *Xenopus* oocyte[6] and active nematic fluctuations in the mitotic spindle[7]. It is generally accepted that the driving force for many of these processes originate from both filament turnover and the relative sliding between molecular motors and cytoskeletal polymers along their long axis[8,9,10]. For example, *in vitro*, rigid microtubule filaments[11,12] reach a flowing dynamic steady state under the influence of kinesin motors[12,13]. As a consequence, microtubule networks retain their overall density and architecture [7,14] or yield to extensile flows[15]. By contrast, myosin motor activity on semi-flexible F (filamentous)-actin leads to filament buckling[16] and severing at high curvatures[17,18]. As a result, F-actin networks experience macroscopic architectural changes[19] and large strains[16] during destabilizing contractile flows[20]. Thus, it remains unclear how networks of semi-flexible polymers can maintain a dynamic steady state in the presence of active stress. More generally, the relationship between out-of-equilibrium activity and the accumulation and dissipation of mechanical stresses in the stabilization of active materials is unknown.

In this work, we characterize the thermodynamic criteria for the maintenance of dynamic stability in an active nematic material composed of semi-flexible F-actin through determination of the rate of entropy production as a function of molecular motor activity. First, we systematically identify the range of motor activity that differentiates macroscopic contractility (unstable) from steady-state non-contractile behavior (stable). Next, we determine the effect of activity on the microscopic balance of mechanical work and the production of entropy from the myosin-induced bending of individual F-actin. This provides a quantitative relationship between how far the system is from equilibrium with its propensity to dissipate mechanical energy. We then correlate network and filament properties to associate the accumulation of mechanical work and the production of entropy with the mechanical stability of the bulk material. Finally, we calculate the entropy produced per individual myosin filament and correlate the motions of myosin filaments with the bulk dissipation that stabilizes the material.

## Results

### Semi-flexible F-actin Self-assembles into a 2D Network with Nematic Ordering

F-actin is crowded to the surface of a phospholipid bilayer over time due to the depletion forces induced by methylcellulose (0.25% MC)[21] (Supplementary Movie 1). In the absence of adhesion between actin filaments and membrane, the filaments change their spatial orientation to establish a net direction upon reaching the membrane surface. This reorganization of the network generates local domains of nematic alignment, quantified by the coarse-grained nematic order parameter $\langle q \rangle = 2 \langle \cos^2 \theta - 1/2 \rangle$ (Fig 1 a-d, Methods, Supplementary Figure 1, Supplementary Note). As F-actin accumulates on the bilayer, the network transitions continuously from an isotropic to a nematic phase (Supplementary Figure 1,2). The nematic domains originate from and terminate in regions of disorganized F-actin containing disclination defects with topological charge $\pm 1/2$ [22]. -1/2 defects are formed by moderate F-actin bending in radial directions around a central void, whereas +1/2 defects form due to highly bent F-actin oriented along a single direction about a central core (Fig 1 e,f).

While the F-actin network exhibits the same defects and symmetries as a traditional nematic liquid crystal composed of short, rigid rods[12, 23], the average F-actin length in our experiments is ~10 μm, comparable to their persistence length[24]. As such, defects form due to bending and entanglement of a small number of individual actin filaments (Fig 1e,f, bottom). Further, we do not observe defect motion or annihilation, reported in other biopolymer nematic liquid crystals[12, 25], thus the effect of myosin activity on network architecture and stability remains unclear.

**Myosin Activity Impacts Nematic Ordering in the Absence of Contractile Flow**

Thus far, the dynamics of myosin-driven F-actin networks have largely been attributed to contractile flow [16, 17, 19, 20, 26, 27, 28, 29, 30]. We have previously shown that contractile flow occurs in a cooperative manner above a critical myosin thick filament density, $\rho_c$[17][31]. Thus, our first metric for motor 'activity' is one that is dependent on myosin density and results in contractile stresses, $\zeta(\rho)$. For $\rho > \rho_c$, filament buckling coincides with network contraction as it shortens the filament end-to-end length[32, 33]. However, the impact of sub-contractile densities of myosin ($\rho < \rho_c$) on the dynamics of F-actin is unclear. To this end, we accumulate myosin thick filaments on the assembled F-actin nematic network at densities above and below $\rho_c$ to compare and contrast the impact of activity on actomyosin network assembly dynamics.

To assemble the actomyosin network, myosin dimers are added to the nematically ordered F-actin at $t=0$, quickly accumulating and forming thick filament assemblies within ~100s. After the addition of myosin, we calculate the divergence of the F-actin velocity field, which yields the macroscopic strain rate, $\psi(t) = \langle \nabla \cdot \vec{v}(t) \rangle$. We also calculate the spatially-averaged nematic order of the F-actin network, $\langle q(t) \rangle$ (Fig 2 a,b, Supplementary Figure 3,4) and the change in the average nematic order $\delta\langle q(t)\rangle$, defined as the difference between the nematic order prior to the addition of myosin to the nematic order at time $t$. For $\rho > \rho_c$, the strain rate decreases during contraction until it reaches a maximum in its magnitude, $\psi_{max}$. Relatedly, the change in the nematic order also increases with time, representing a loss in F-actin alignment due to myosin activity. However, the loss of $\langle q(t) \rangle$ precedes the drop in $\psi(t)$ in time (Supplementary Movie 2), suggesting there may be dynamics of actomyosin ($\delta q \neq 0$) that are non-contractile ($\psi \sim 0$).

Indeed, for $\rho < \rho_c$, $\psi_{max}$ decays to zero although $\delta q$ does not – the network retains up to 70% of its original nematic order (Fig 2 c,d, Supplementary Movie 3). Thus, myosin activity can drive the establishment of a steady state defined by changes in structural dynamics absent of contractile flows. To this end, we characterize the dynamics of this steady state by analyzing the fluctuations in the nematic alignment of the F-actin network using both experiment and simulation.

**Active Nematic Fluctuations Are Elevated in Non-Contractile Actomyosin**

Kymographs of F-actin fluorescence intensity exhibit distinct fluctuations in space for thermal (T) and stable networks (S) (Fig 3 a,b). To quantify these fluctuations, we measure the Fourier transform of the equal-time spatial autocorrelation function of the F-actin density fluctuations ($S_{\rho\rho}$) and the nematic director fluctuations ($S_{nn}$) perpendicular to the local axis of F-actin alignment as a function of wavenumber $k_\perp$ (Fig 3 c,d, Supplementary Note, Supplementary Figure 5).

The addition of myosin motors ($\rho < \rho_c$; S) increases the magnitude of the fluctuation present as an elevated $S_{\rho\rho}$ at low $k_\perp$, consistent with the qualitatively larger fluctuations in F-actin fluorescence intensity in kymographs (Fig 3 a,b). Using a model for active nematic gels, this is indicative of an increase in filament mobility or orientational noise (Supplementary Note). While the active gel model fits for short, rigid filaments ($\bar{l} \approx 2.5$ μm, $N_{fils} = 200$), we find deviations from this model for long filaments at high $k_\perp$, suggesting the relationship between activity and filament fluctuations is complex (Supplementary Figure 6). Using agent-based simulations[34], we show that this deviation at high $k_\perp$ arises from length dependent but activity independent dissipative effects (Supplementary Figure 7).

Therefore, a uniform active stress and simple viscosity may not fully capture the dependence of myosin activity on F-actin fluctuations and dissipation. To this end, we investigate the role of motor activity on dissipation on the filament scale by using a model-free estimate of the rate of entropy production.

**Activity-Dependent Dissipation is Maximized in the Stable State**

When driven out of equilibrium, microscopic systems obey fluctuation theorems that relate the irreversibility of a process to the amount of entropy produced in that process[35, 36, 37]. Myosin motors operate far from equilibrium by hydrolyzing ATP to generate forces on F-actin, as previously quantified by deviations from the fluctuation dissipation theorem[38]. Here, for the first time, we show that the dissipated energy in actomyosin is amenable to quantification using the framework of stochastic thermodynamics[37].

Using experiments where only 2% of filaments are fluorescently labeled, individual filaments are tracked over time (Fig 4a). Filament shape is specified by its tangent angle along its arc length at each time, $\theta(s,t)$. This function is decomposed into a set of orthogonal bending modes [39], $\theta(s,t) = \Sigma_q a_q(t) f_q(s)$. Filament dynamics are represented by the trajectory of a point in a phase space spanned by the mode coefficients, $\vec{a}(t) = (a_1(t), a_2(t), ...)$ (Supplementary Figure 8). Such phase space trajectories have recently been used to identify broken detailed balance in mesoscopic biological systems[40, 41, 42] and to calculate rates of energy dissipation in open biochemical systems[43, 44]. We use the phase space trajectories to quantify the entropy produced over time using a formulation based on a Langevin equation for the bending modes[45]. Using natural units, the total entropy produced up to a time $t$ is given by

$$\Delta S(t) = \int_0^t d\tau\, \dot{\vec{a}}^T(\tau)\, \mathbf{D}^{-1}\, \vec{v}^{ss}[\vec{a}(\tau)] \qquad (1)$$

where $\vec{v}^{ss}[\vec{a}(\tau)]$ is the steady state phase space velocity estimated using the entire trajectory, and $\dot{\vec{a}}(\tau)$ is the instantaneous phase space velocity. **D** is the diffusion matrix that enters into the Fokker-Planck equation associated with the underlying Langevin equation. For simplicity, we estimate **D** from the drag coefficients of a slender rod[46]. We verify our calculations and approximations by checking that a control system obeys the detailed fluctuation theorem (See Supplementary Note for details, Supplementary Figure 9).

Using the above formalism, we calculate the average total energy dissipated per unit filament length as $\overline{\Delta s}(t)T$, where $\Delta s(t)=\Delta S(t)/L$, $L$ is the filament length, $T$ is the temperature of the surrounding medium, and the bar denotes an ensemble average taken at each time point. We see that the actomyosin system shows three distinct phases of energy dissipation that correspond to the three states of actomyosin discussed above (Fig 2d). In these experiments, myosin accumulates over time, and the three states coincide with changes in the number of tracked myosin thick filaments over time (Methods). The first state is a passive state distinct from thermal states due to the presence of myosin dimers that have not yet formed myosin thick filaments ($S_0$). The second is an active, non-contractile state as myosin thick filaments begin to appear ($S_1$). The third is the contractile state (C) where myosin thick filaments begin to aggregate and thus the number of tracked thick filaments decreases (Fig 4b, black). State $S_0$ shows a small increase in energy dissipated, followed by a large increase in the rate of energy dissipation during state $S_1$. State C shows a decreased energy dissipation rate (Fig 4 b,c). Within state $S_1$, total energy dissipation as a function of myosin number density across experiments collapses along a single curve until the system gets closer to the contractile regime, indicating that myosin filaments dissipate energy uniformly below a number density, beyond which they do not behave identically (Fig 4d). These results are replicated using agent-based simulations (Fig 4 b-d, insets, Supplementary Movie 4,5).

Having investigated the role that activity plays in dissipating mechanical energy, we next sought to understand how activity also stores mechanical energy in the system via filament bending. To this end, we measured the change in bending energy per unit filament length as myosin accumulates, $\overline{\Delta \varepsilon_{bend}} = \overline{\Delta E_{bend}/L}$ (Fig 4e, Methods). As with the dissipation energy, we take the ensemble average across filaments at each time point. We again see three distinct regimes, where the actin bending energy does not change during state $S_0$ and increases rapidly in state $S_1$ (Fig 4f). In state C, bending energies are elevated but decreasing nominally. This may be attributed to filament severing in experiment[16], although it is not necessary as simulations without severing show similar results (Fig 4 e,f, insets, Supplementary Figure 10). In simulation, it can be observed that upon the cessation of contractile flow, filaments are polarity sorted with motors at filament barbed ends and bends are released (Supplementary Movie 6). Again, plotting bending energy as a function of myosin number density in state $S_1$ collapses experiments along a single curve as the system approaches contractility (Fig 4g).

Thus, the non-contractile state (S) dissipates the most energy as measured through non-equilibrium entropy production. As the stable (S) and contractile states (C) have different entropy production rates, we sought to determine if there was a difference in the underlying actomyosin interactions that produce these rates.

**Diversity in Actomyosin Motions Underlies Dissipation in Stable State**

Active transverse fluctuations and F-actin bending may suggest that myosin and F-actin are not aligned, in contrast to the canonical model for their interaction is anti-parallel filament sliding. Here, we quantify the extent of axial vs perpendicular actin motions and compare them to myosin motions.

To quantify the extent to which non-contractile networks exhibit perpendicular bending motions, we measure the anisotropic velocity autocorrelation, defined as $\delta C_{vv}(r) \equiv \langle C_{vv}^{\perp}(r,t)/C_{vv}^{\perp}(0,t) - C_{vv}^{\parallel}(r,t)/C_{vv}^{\parallel}(0,t)\rangle_t$ (Supplementary Note). Positive values indicate enhanced perpendicular fluctuation autocorrelations; negative values indicate enhanced parallel fluctuation autocorrelations. We find that all stable systems, regardless of myosin isoform, exhibit greater fluctuations perpendicular to the filament axis, in stark contrast with contractile systems that show larger autocorrelations parallel to the filament's axis as would be expected for sarcomeric contraction[8, 47] (Fig 5 a,b). We name these reversible, myosin-derived transverse fluctuations "plucking" (Supplementary Movie 7).

Using a light activation assay[31] with skeletal muscle myosin II (SkMM), we induce contractility at a constant myosin density. 405 nm light inactivates blebbistatin, an ATPase inhibitor, thereby activating myosin in the area of illumination. During contraction, we measure the relative angle between SkMM thick filaments and the actin it decorates (Fig 5c). The extent of contractility is measured by an increase in $\psi$ (Fig 2). We find that as the magnitude of $\psi$ increases, we see a rapid change in the relative angle ($\delta\theta$) between actin and myosin (Fig 5d). We therefore attribute the enhanced perpendicular fluctuations of actin in a non-contractile state, and therefore the enhanced entropy production rate, to the variation in the relative angles between individual myosin assemblies and F-actin. Thus, we append our metric for activity to include both the density of thick filaments $\rho$, but also a wider spectrum of actomyosin interactions which is quantified by the relative angle between motors and filaments, $\theta$, i.e. $\zeta(\rho) \to \tilde{\zeta}(\theta, \rho)$ (Fig 5e)

**Discussion**

Through the engineering of a novel active nematic, we identify a previously unobserved phase of actomyosin – a structurally dynamic state, absent of contractile flow or filament turnover at an intermediate level of activity. These dynamics include myosin-driven fluctuations in both the nematic director and F-actin density. Active nematic theory provides a general framework for understanding the coupling between these fluctuations. However, we find that the model does not fit at all length-scales. For low $k_\perp$, we find dramatically increased density autocorrelations due to myosin activity, indicative of an increase in F-actin mobility. A poor fit to the data challenges the applicability of the active gel model, for two possible reasons. First, the active stress $\zeta$ assumes that forces are applied along the axis of alignment. It was not clear *a priori*, that this motion that is assumed to drive contractility also occurs in a dynamic steady state. Second, the dissipation represented by a simple viscosity $\eta$, may in fact be activity dependent. Thus, we seek to calculate the effect of activity on dissipation in a model independent way, which makes no assumptions on the specific functional form of the active stress.

Entropy production rates provide a measure for how far a system is from equilibrium[36, 48] and the extent to which mechanical energy is dissipated. Unexpectedly, we find that the rate of entropy production is non-monotonic with increased activity. As the system is driven from equilibrium, the dissipation rate first increases in the stable state and then attenuates in the contractile state. In addition, the work applied, as indicated by the filament bending energy, increases in then decreases due to relaxations via polarity sorting and filament severing (Supplementary Figure 10, Supplementary Movie 6). Thus, while the contractile state has the highest entropy, it is the stable state in which the rate of entropy production is maximized (Fig 5e).

Finally, while we find that axial motions of F-actin, consistent with the canonical "sliding" of F-actin in muscle, are associated with contractility, stable fluctuations are surprisingly dominated by transverse filament deformations (Supplementary Movie 7), regardless of myosin isoform, indicating its generality. These reversible F-actin plucking events arise from transient and diverse interactions between non-aligned myosin and F-actin (Fig 5). A lack of alignment or overlap between myosin and actin filaments would imply that fewer myosin heads may be involved in the generation of stress. Indeed, we find that only a few heads are sufficient to induce the observed bending energies (Supplementary Figure 2). Thus, this work challenges the prevailing model of molecular motors as purely idealized force dipoles whose relative motions with F-actin always yield to contractility. Likewise, the definition of motor-based activity is now more complex; there is a spectrum of interactions that occur in disordered assemblies of myosin and F-actin at the molecular level. That spectrum in turn, may determine system level entropy production and dissipation that stabilizes active materials.

**Conclusion**

The relationship between motor activity and the accumulation and dissipation of mechanical energy, which determines material stability is complex. The complexity arises from the diversity of motor-filament interactions, and the specific impact those interactions have on how far the material is pushed from equilibrium and the rate at which the entropy is produced. This multi-length scale identification and characterization of active stability presents a new and comprehensive understanding for the dynamics of active biological materials.

**Acknowledgements**

We thank Enrique de la Cruz for help with reagents and supplies. We acknowledge funding CMMI-1525316, ARO MURI W911NF-14-1-0403, NIH RO1 GM126256 and NIH U54 CA209992 to MM. SB acknowledges Strategic Fellowship support from the University College London. IAL acknowledges support from NSF Fellowship grant # GE-1256259. DSS acknowledges support from NSF Fellowship grant # DGE1122492. Any opinion, findings, and conclusions or recommendations expressed in this material are those of the authors(s) and do not necessarily reflect the views of the National Science Foundation. We thank Benjamin B. Machta, Pierre Ronceray, and Francois Nedelec for helpful discussions.

**Author Contributions**

MPM and SB designed and conceived the research. MPM, IL, DSS, AT, VY, APT, and SB acquired experimental data and performed analysis. MPM, AT, DSS, SB, VY, PO, and IL contributed analytical methods for experimental data. VY performed computational simulations. MPM & DSS wrote the paper.

**Competing Financial Interests**

The authors declare no competing financial interests.

## Methods

**Assay construction**

Assay Construction is performed as described previously[49]. Briefly, a phospholipid bilayer composed of 99.8% Egg Phosphatidyl-Choline (EPC, Avanti Polar Lipids) and 0.2% Oregon Green 1,2-Dihexadecanoyl-sn-Glycero-3-Phosphoethanolamine (OG-DOPC, Molecular Probes) is deposited onto a UV-treated or Piranha-treated coverslip. Stabilized F-actin (2 μM) with 0.3 μM fluorescent actin (Alexa-Fluor 568, Molecular Probes) is crowded to the surface of the bilayer using 0.25% methyl-cellulose (Sigma). Skeletal-muscle myosin (10.5 μM stock), smooth-muscle myosin and non-muscle myosin II monomers (2.4 μM stock) are then added to the F-actin network. As the actual concentration within a microscopic field of view can vary, the precise concentration of myosin is calculated by visualizing the labelled myosin in experiment, and calculating the density of myosin thick filaments per unit area (#/μm$^2$) from the fluorescence images[31, 50, 51].

**Myosin isoforms**

Smooth muscle and non-muscle isoforms are purified, polymerized and fluorescently labelled according to standard protocols[49]. Skeletal muscle myosin (Cytoskeleton) and other isoforms are stained using Alexa-647 (Molecular Probes). Skeletal muscle myosin is used to assess alignment between thick and thin filaments due to its length (~2 μm). Smooth muscle myosin is used to assess F-actin deformation due to slower motor kinetics. Skeletal and non-muscle isoforms are also used for light activation as they are sensitive to blebbistatin ATPase inactivation.

Confocal microscopy is used to image the dynamics of fluorescent protein with the assay. Images are recorded on a Zyla 4.2 Megapixel sCMOS camera on a Leica microscope and also recorded on a Coolsnap Hq2 CCD Camera on a Nikon Ti Inverted Microscope.

**Nematic order parameter calculations**

The nematic order parameter, $q$, is calculated using custom Matlab code. First, a director field is created from images of fluorescently labeled F-actin[52]. Briefly, fluorescent images are divided into small, overlapping 3.5 μm by 3.5 μm windows, and the local F-actin orientation (director) is calculated for each window, yielding an F-actin director field over an image. To determine the local F-actin director, each window is Gaussian filtered and transformed into Fourier space using a 2D fast Fourier Transform (FFT). The axis of the least second moment was calculated from the second order central moments of the transformed window, and the angle of the local F-actin director is defined as orthogonal to this axis. Next, the local degree of alignment is calculated between adjacent windows within 3x3 kernels. The local nematic order is calculated for the central window in each kernel using the modified order parameter equation $\langle q \rangle = 2 \langle \cos^2 \theta - 1/2 \rangle$, where $\theta$ is the difference in F-actin orientation between the central window and the 8 surrounding windows. This process is repeated for all possible 3x3 kernels over an image, yielding a nematic director field with defined director magnitude and orientation for each window over an image. Perfect alignment between adjacent regions within an F-actin network results in an order parameter equal to one. Conversely, orientation differences of 45 degrees (maximum expected for quasi-2D F-actin network) between adjacent regions of the network result in an order parameter equal to zero.

**Flow calculations**

Particle image velocimetry (PIV) is applied in Matlab (mPIV, https://www.mn.uio.no/math/english/people/aca/jks/matpiv/) to fluorescent F-actin images, yielding displacement & velocity vector fields.

**Fluctuation autocorrelations**

Using custom Matlab code, the local alignment field, nematic order parameter, and density scalar field are extracted for each confocal image. The equal time autocorrelation function for both the nematic director and the density fields are calculated at every position in the image along an axis transverse to the local alignment field, and the results are binned across all time and space for each experimental condition. See Supplementary Note for more details.

**Agent-based simulations**

The experiments are simulated in an open source package, Cytosim[34]. Actin filaments are modeled as polar worm like chains composed of rigid segments of length 0.1 μm. The simulation volume is quasi-3D with periodic boundary conditions along x and y axis to limit finite size effect. The boundaries along z-axis are closed to mimic the action of methyl cellulose. The thickness of simulation volume was set to twice the diameter of an actin filament. The filaments initially grow for a fixed period of time to reach a predetermined length distribution and volume fraction and then relax to reach an equilibrium steady state. The motors are modeled as Hookean springs that can bind with filaments and move along them towards their barbed end. The details regarding implementation of simulation are described in Supplementary Note.

**Calculating entropy from experimental data**

In order to calculate the entropy produced by individual filaments, we first track individual filaments using the ImageJ plugin JFilament[53]. Using custom MATLAB scripts, we decompose the filament shapes into a set of orthogonal bending modes. The coefficients for each bending mode are then tracked over time in the configurational phase space spanned by the mode coefficients. We coarse-grain the resulting velocity field in phase space[42] in order to obtain a steady-state velocity to be used in Eq. (1) to calculate the entropy at each time point for every filament, dividing by the filament length to create an intensive variable. The mean entropy across all filaments present at a given time point is taken, and the cumulative sum of this mean is plotted in Fig 4e. See Supplementary Note for more details.

**Calculating filament bending energy**

Filaments are tracked in time as mentioned above and represented as $M$ points along the filament. A circle is fit to rolling sections of $n \ll M$ points, and the inverse of the radius of the resulting circle is taken as the local curvature of the filament at that point, $\kappa$. The bending energy of the entire filament is then given by $E_{bend} = \frac{EI}{2} \int_0^L \kappa^2(s)\, ds$. $EI$ is the flexural rigidity of actin[54] The ends of the filaments are precluded from this measure because points within $n/2$ points of the end cannot have a set of $n$ points centered around them.

**Figures**

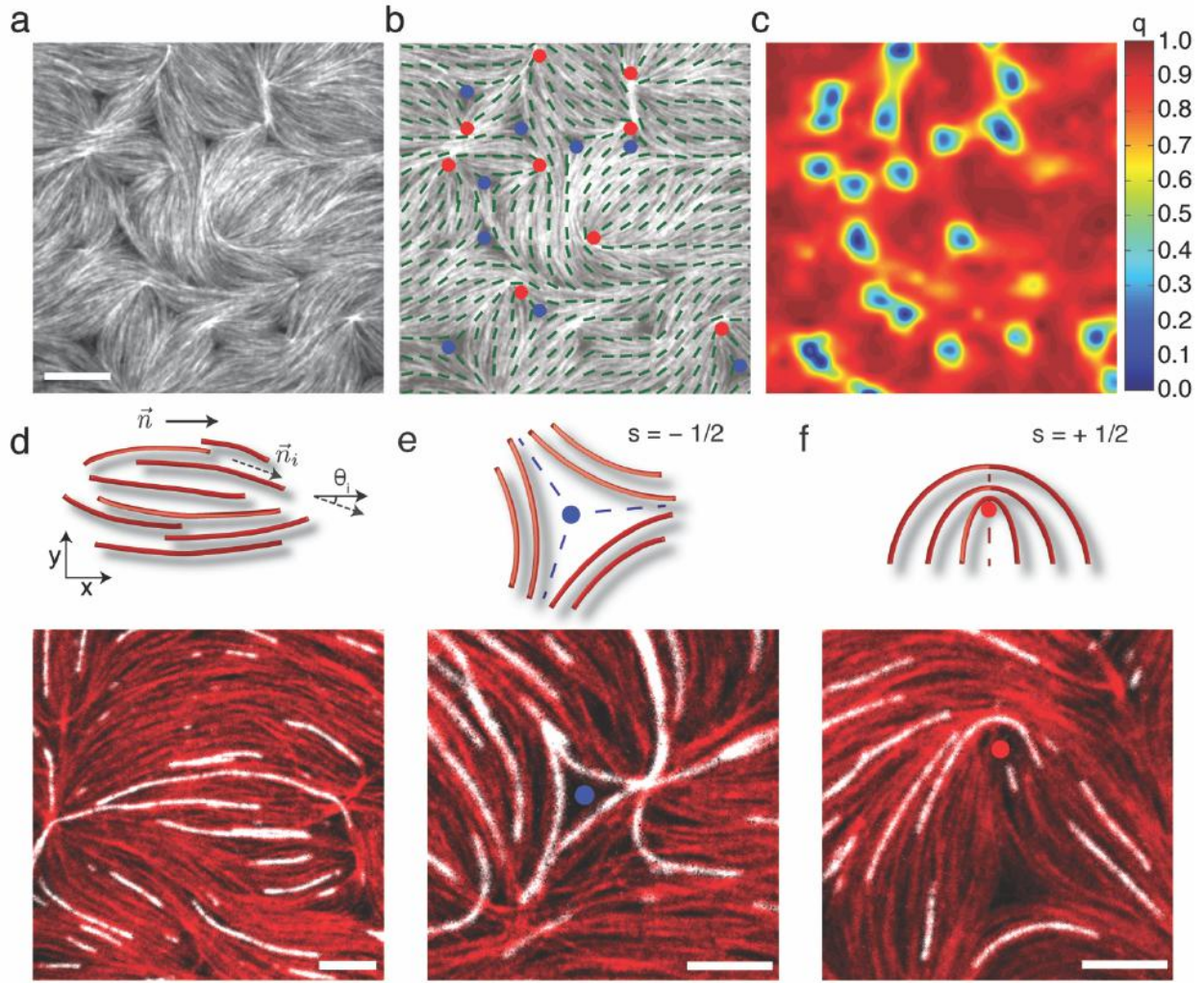

**Figure 1: Synthesis of a 2D actomyosin network with nematic ordering**

(a) Fluorescent F-actin crowded to the bilayer. Scale bar is 10 μm. (b) Location of disclinations with topological defect charges s=-1/2 and s=+1/2, shown by blue and red dots respectively. Green lines represent nematic director of the F-actin gel. (c) Heat map of the nematic order parameter $q = 2 \langle \cos^2 \theta - 1/2 \rangle$. (d) (top) Schematic of a nematically ordered domain comprised of many actin filaments, where $\vec{n}$ is the nematic director of the entire domain, $\vec{n}_i$ is the local alignment of a single F-actin, and $\theta_i$ is the angle between them and (bottom) image of a single nematic domain.

Both red and white show actin filaments labeled with different fluorophores, polymerized separately, and combined at a 1:50 ratio prior to crowding to visualize individual filaments within the larger network. (e) Schematic (top) and image (bottom) of a -1/2 disclination topological defect and local nematic ordering in quasi 2D F-actin. (f) Schematic (top) and image (bottom) of a +1/2 disclination topological defect and local nematic ordering in quasi 2D F-actin network. Scale bars in d-f are 5 μm.

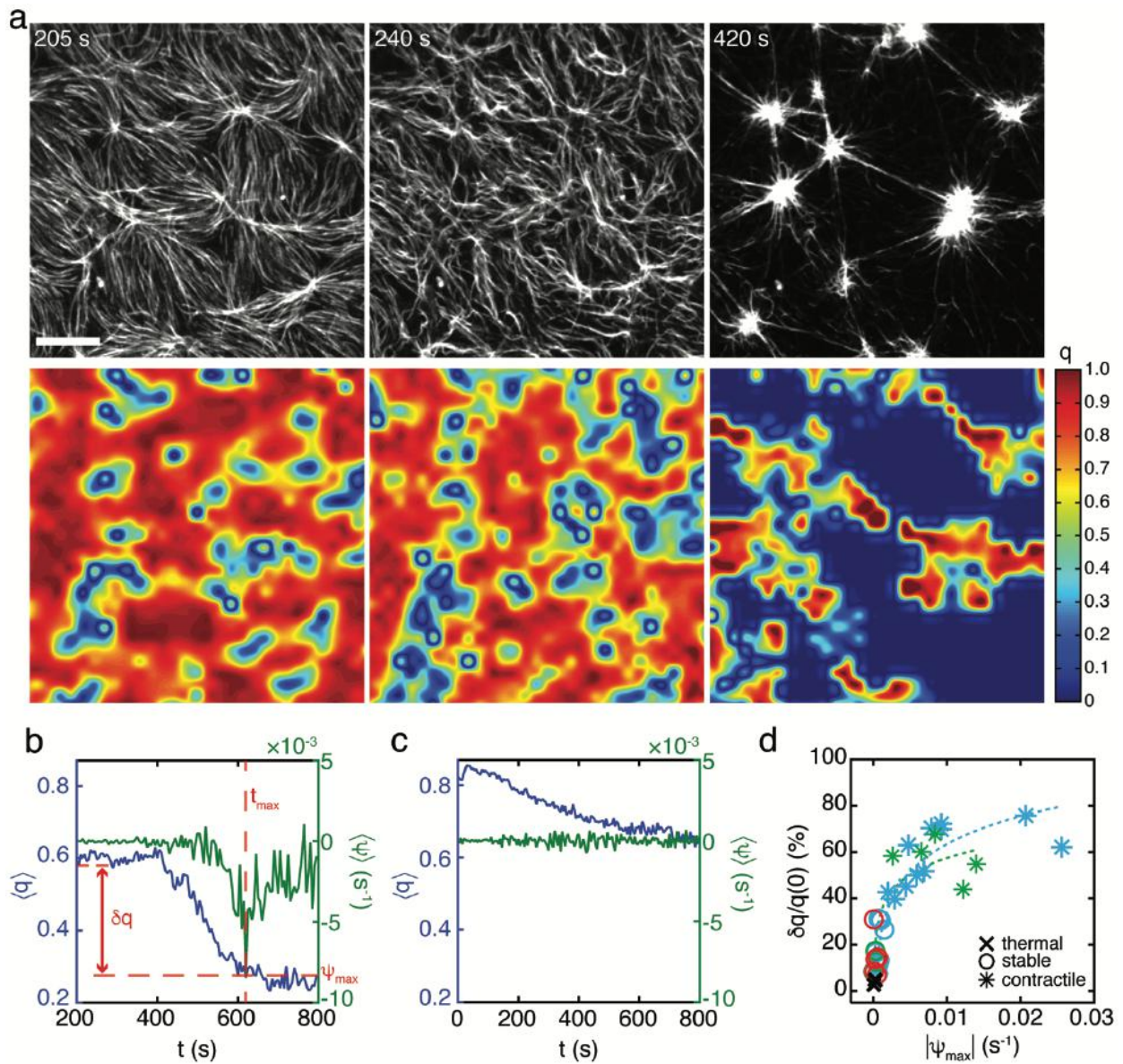

**Figure 2: F-actin nematic order is altered by myosin density in both contractile and stable states**

(a) (top) Fluorescent F-actin network undergoing contraction. Scale bar is 10 μm.. (bottom) Heat map of scalar nematic order parameter $q$. Myosin dimer added at $t = 0$. (b) Spatially averaged F-actin nematic order parameter measured ($q$, blue) and divergence of F-actin velocity ($\psi$, green) 200s after the onset of myosin addition when $\rho > \rho_c$. Time of maximum magnitude of divergence ($t_{max}$) indicated by vertical dotted red line. Difference between nematic order at $t=0$ and at time of maximum divergence ($\delta q = q(0) - q(t_{max})$) indicated by horizontal dotted red line. (c) Spatially averaged nematic order (blue) and divergence of the velocity (green) for a stable actomyosin network, $\rho < \rho_c$, where myosin is added at $t = 0$. (d) Percent change in nematic order ($\delta q/q(0)$) for thermal (×), stable (○), and contractile (*) network states. The marker color denotes the myosin isoform added to each experiment (SkMM= blue, SmMM = green, NMM = red, no myosin = black). We define non-contractile, 'stable' (S) networks as those with $\psi_{max} < \psi_c = 2 \times 10^{-3}$ s$^{-1}$, contractile networks (C) for $\psi_{max} > \psi_c$, and networks for which no myosin added as 'thermal' networks (T).

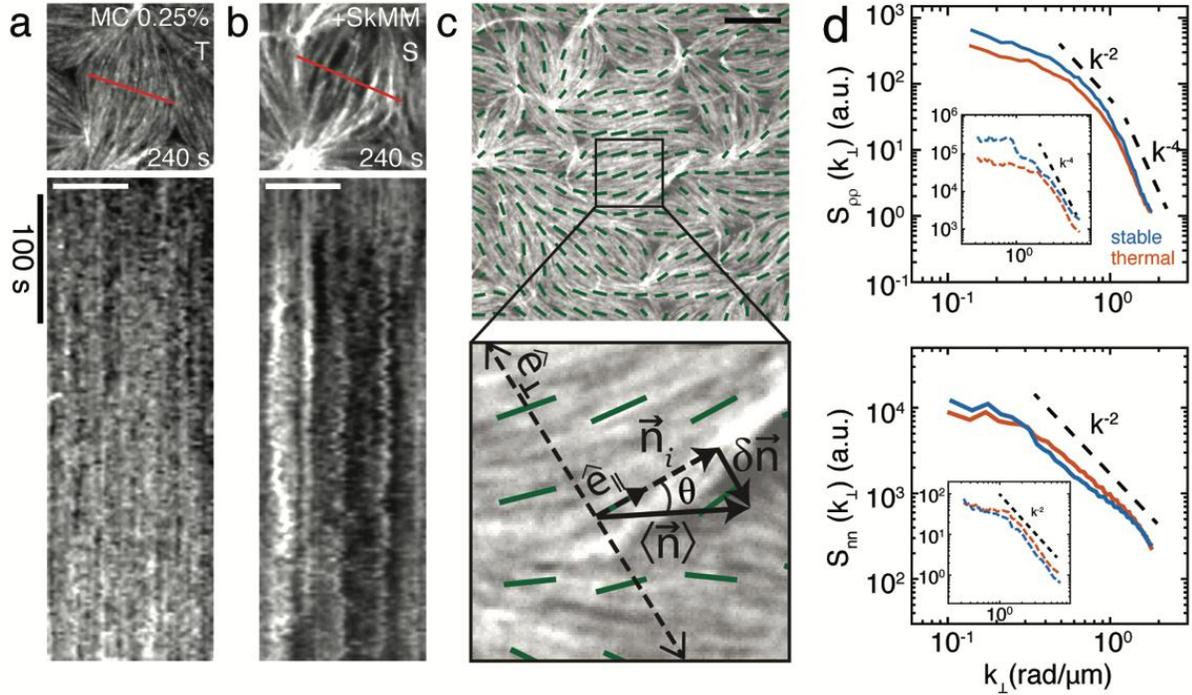

**Figure 3: Intermediate activity enhances fluctuations in stable actomyosin**

(a,b) Fluorescent F-actin kymograph across a single nematic domain of a 2D network in the presence of 0.25% MC (T, a), and 0.25% MC + SkMM (S, b). White scale bar is 5 μm (c) 2D F-actin network in the presence of 0.25% MC with alignment vector field (green), scaled by the local nematic order parameter, overlaid. Lower panel shows a schematic of the local alignment vector ($\vec{n}_i$), local coordinate system defined by the axes parallel ($\hat{e}_\parallel$) and perpendicular ($\hat{e}_\perp$) to the local alignment vector, and local alignment vector fluctuations ($\delta\vec{n}_i \equiv \vec{n}_i - \langle\vec{n}_i\rangle$) used to calculate fluctuation autocorrelations. Scale bar is 10 μm. (d) Equal time spatial density-density ($S_{\rho\rho}$, top) and director-director ($S_{nn}$, bottom) fluctuation autocorrelations as a function of the perpendicular wave vector ($k_\perp$). The spatial autocorrelations are shown for thermal (orange, 0.25% MC, N = 16) and stable (blue, SkMM, N = 6) experimental conditions within the F-actin network. All experiments contain the same concentration of F-actin (2.32 μM) with 0.25% MC. The dashed black lines follow the predicted autocorrelation decay by active nematic liquid crystal theory. Insets show corresponding results for $S_{\rho\rho}$ and $S_{nn}$ from agent-based simulation. Each result is an average over 3 simulations. Mean length of simulated filaments is 4.3 μm.

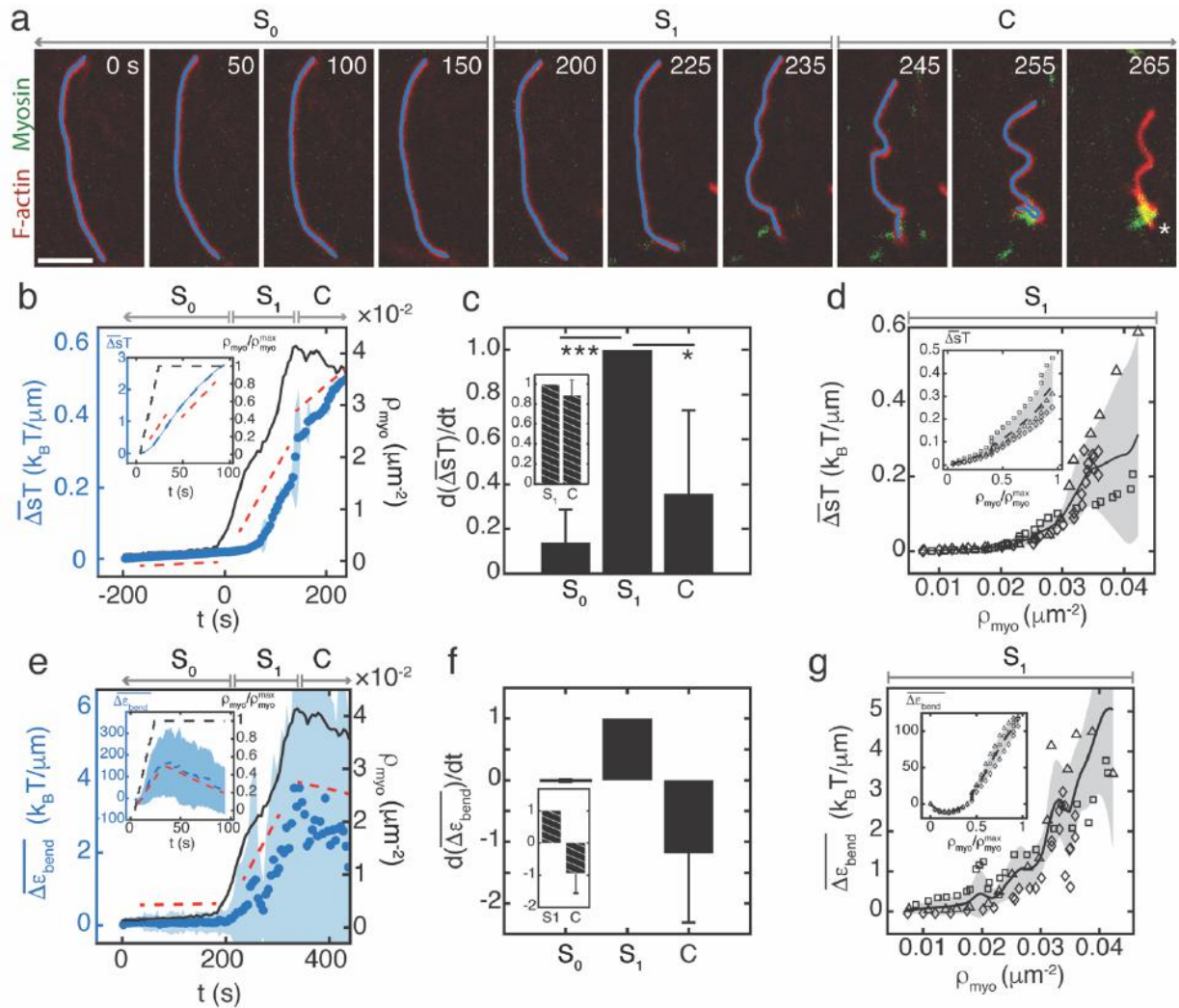

**Figure 4: Entropy production rate is highest in stable actomyosin**

(a) Example of experiment with 1% labeled filaments (red) as myosin accumulates (green). Filaments are tracked (blue line) until a severing event, indicated by white asterisk. Gray arrows indicate three states of entropy production rate: stable prior to myosin thick filament formation ($S_0$), stable as myosin thick filaments accumulate ($S_1$), and contractile (C). Scale bar is 4 μm. (b) Ensemble averaged energy dissipated per unit length, $\overline{\Delta sT}$, as a function of time (blue) and number density of myosin thick filaments counted as a function of time (black). Blue dots and shaded areas are mean ± standard deviation of n=19 filaments tracked in a single experiment. Experiment is broken into three phases, $S_0$, $S_1$, and C. Red dashed lines indicate slopes measured in each state. (c) Means ± standard deviation for slopes of entropy in states $S_0$, $S_1$, and C for n=4 experiments. Each experiment's slope is normalized to the slope of $S_1$ in that experiment. $p<10^{-4}$ between slopes $S_0$ and $S_1$, and p=0.014 between slopes $S_1$ and C. (d) Dissipation energy density as a function of myosin number density in state $S_1$ for n=3 experiments, indicated by different symbols. Black line and shaded area are mean ± standard deviation across experiments. (e) Similar to (b), but showing filament bending energy per unit length, $\overline{\Delta\varepsilon_{bend}}$, in blue. (f) Similar to (c), but for filament bending energy slopes. $p<10^{-5}$ between slopes $S_0$ and $S_1$, and p=0.001 between slopes $S_1$ and C. (g) Similar to (d) but for filament bending energy. All insets for b-g show recapitulation of data in main figure by agent-based simulations for n=3 simulations. In the simulations, there is no $S_0$ phase because myosin is added immediately at $t=0$.

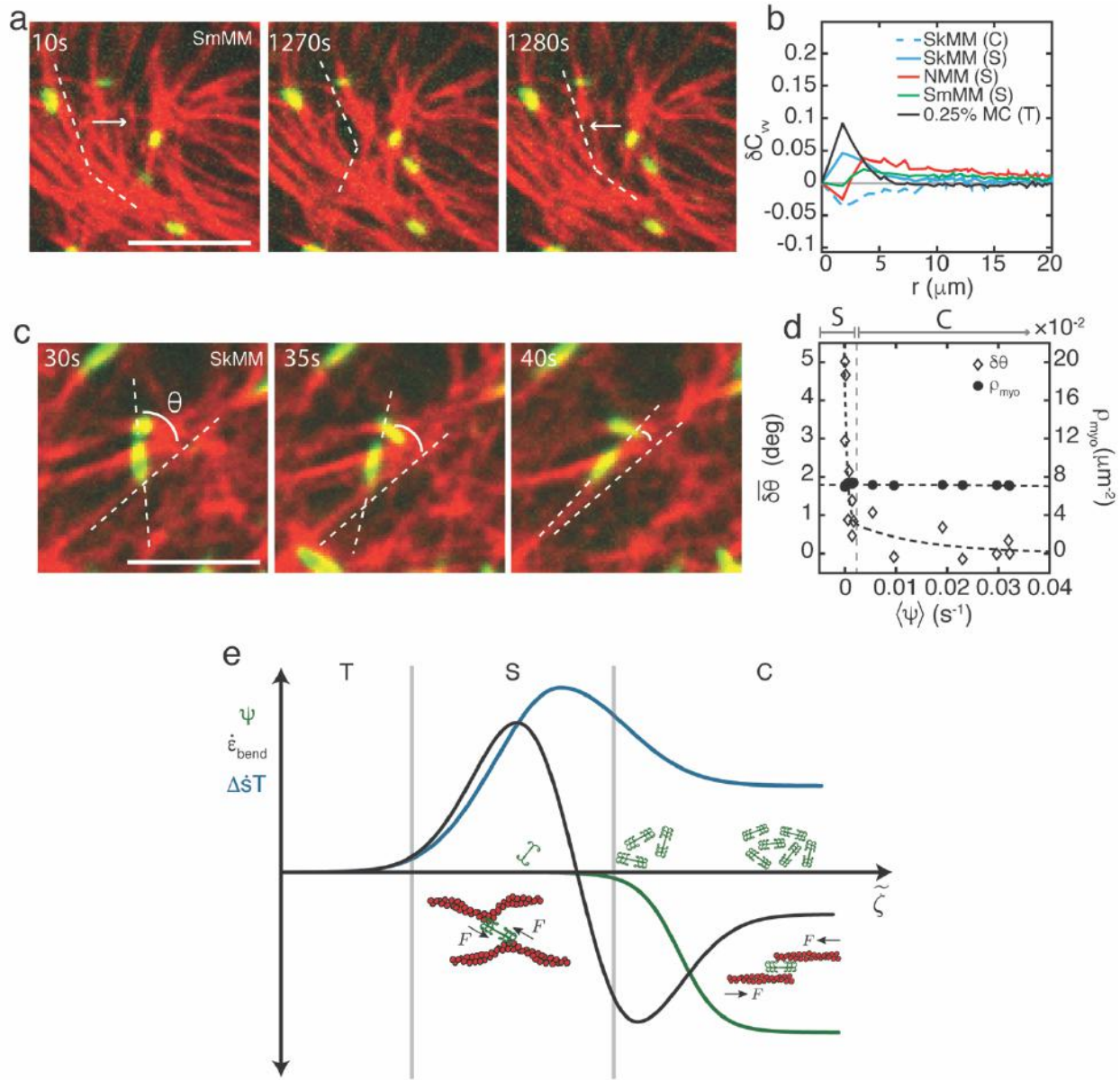

**Figure 5: Misaligned myosin bends F-actin perpendicularly in stable actomyosin**

(a) Motions of F-actin within an actomyosin network. Images of 40 nM smooth muscle myosin (green) embedded within ~ 2 μM F-actin network (red). White dotted lines indicate alignment of F-actin. White arrows indicate direction of motion of F-actin. Scale bars are 5 μm. (b) Anisotropic velocity-velocity autocorrelation, $\delta C_{vv}$. Averages are taken across several experiments ($N_{SkMM}$ = $N_{0.25\%MC}$ = $N_{0.15\%MC}$ = 3, $N_{SmMM}$= 5, $N_{NMM}$ =2), where each experiment is represented by its own temporal average. The colors represent different experimental conditions. (c) Images of ~2 μM actin and 40 nM skeletal muscle myosin. Angle $\theta$ is measured between the myosin thick filament and the underlying F-actin. Scale bar is 5 μm. (d) Ensemble average change in angle $\overline{\delta\theta(t)} = \overline{\theta(t) - \theta(\infty)}$ of myosin thick filaments (open diamonds) and mean myosin number density (filled dots) as functions of $\psi$, which increases from 0 to 0.04 s$^{-1}$ as time increases. Dotted lines are guides for the eye. Stable (S) and contractile (C) states indicated above plot. (e) Schematic phase diagram showing how dissipation rate (blue), bending energy rate (black), and $\psi$ (green) all change as activity, $\tilde{\zeta}$, increases. An increase in $\tilde{\zeta}$ coincides with an increasing myosin density, indicated by the green myosin cartoons. Thermal, stable, and contractile states are indicated by T, S, and C respectively. In the stable state, a schematic representation of myosin's perpendicular effects on F-actin are shown. As $\tilde{\zeta}$ increases, myosin is shown sliding anti-parallel filaments.